\documentclass[12pt]{article}
\pdfoutput=1
\usepackage{amsmath,amssymb,amsfonts}
\usepackage{a4wide,epsfig,psfrag,scalefnt}
\usepackage[dvipsnames]{xcolor}
\usepackage{braket}
\usepackage{placeins}
\usepackage{breqn}
\usepackage{slashed}
\usepackage{enumitem}
\usepackage[numbers,sort&compress]{natbib}
\usepackage{caption}
\usepackage{subcaption}
\usepackage{soul}

\graphicspath{ {figs/} }

\allowdisplaybreaks

\begin{document}

\title{\vskip-3cm{\baselineskip14pt
    \begin{flushleft}
     \normalsize P3H-23-051, TTP23-029, ZU-TH 40/23
    \end{flushleft}} \vskip1.5cm

  Next-to-leading order electroweak corrections to $gg\to HH$ and $gg\to gH$
  in the large-$m_t$ limit
}
 
\author{
  Joshua Davies$^{a}$,
  Kay Sch\"onwald$^{b}$,
  Matthias Steinhauser$^{c}$,
  Hantian Zhang$^{c}$,
  \\
  {\small\it (a) Department of Physics and Astronomy, University of Sussex,
    Brighton BN1 9QH, UK}
  \\
  {\small\it (b) Physik-Institut, Universit\"at Z\"urich, Winterthurerstrasse 190,}\\
  {\small\it 8057 Z\"urich, Switzerland}
  \\
  {\small\it (c) Institut f{\"u}r Theoretische Teilchenphysik,
    Karlsruhe Institute of Technology (KIT),}\\
  {\small\it Wolfgang-Gaede Strasse 1, 76128 Karlsruhe, Germany}
}

\date{}

\maketitle

\thispagestyle{empty}

\begin{abstract}

  We compute two-loop electroweak corrections to Higgs boson pair and Higgs
  plus jet production, taking into account all sectors of the Standard Model.
  All diagrams with virtual top quarks are computed in an expansion for large
  top quark mass up to order $1/m_t^8$ or more.  We present analytic results for the
  form factors and discuss the convergence properties.
  For the process $gg\to gH$ we also consider QCD corrections in the
  large-$m_t$ limit.

\end{abstract}

%- }}}

\newpage

%- {{{ Introduction:

\section{Introduction}

Higgs boson pair production is a crucial process to obtain deeper insight into
the symmetry breaking mechanism of the Standard Model (SM).  For this reason,
it is one of the most important processes studied in detail at the Large
Hadron Collider at CERN and similarly at its High Luminosity upgrade which
will begin operation within this decade.  The main SM production mechanism for
Higgs boson pairs is via gluon-gluon fusion and a number of higher-order
corrections have been computed, mainly in the context of QCD.  As far as
electroweak corrections are concerned comparatively very little is known.
First steps have been taken in
Refs.~\cite{Davies:2022ram,Muhlleitner:2022ijf}.  In
Ref.~\cite{Davies:2022ram} the two-loop box diagrams have been considered
where a Higgs boson is exchanged between the massive top quarks. It has been
shown that a deep expansion in the high-energy limit leads to results for the
form factors which are valid in a large part of the phase space.  In
Ref.~\cite{Muhlleitner:2022ijf} top-quark Yukawa corrections have been
considered, partly in the infinite top quark mass limit.  
Electroweak corrections proportional to the Higgs self-couplings
have been considered in Ref.~\cite{Borowka:2018pxx}
using a numerical approach.
In the present work
we compute the complete NLO electroweak corrections as an expansion in the
large top quark mass limit, including sub-leading terms up to $1/m_t^{10}$.
The corresponding corrections in the case of QCD have been computed in
Refs.~\cite{Dawson:1998py,Grigo:2013rya,Degrassi:2016vss}.

A similarly important process at the LHC is the production of a Higgs boson in
association with a jet. As for Higgs boson pair production the dominant
production channel is gluon-gluon fusion, with the partonic process $gg\to gH$.
NLO QCD corrections have been considered in a number of works: in the
large-$m_t$ limit \cite{Neumann:2016dny}, in the high-energy limit
\cite{Melnikov:2016qoc,Kudashkin:2017skd,Lindert:2018iug} and numerically,
including exact dependence on $m_t$
\cite{Jones:2018hbb,Chen:2021azt,Bonciani:2022jmb}.  NNLO corrections have
even been computed in the infinite top quark mass
limit~\cite{Boughezal:2013uia,Chen:2014gva,Boughezal:2015dra,Chen:2016zka,Gehrmann:2023etk}.
NLO electroweak corrections via massless bottom quark loops have been computed
in Ref.~\cite{Bonetti:2020hqh}, and the corrections induced by a trilinear
Higgs coupling in the large top mass limit have been recently calculated in
Ref.~\cite{Gao:2023bll}.  In this work we compute, for the first time, the
full NLO electroweak corrections involving virtual top quark loops. We
consider all sectors of the Standard Model and perform an expansion for large
$m_t$ up to order $1/m_t^8$.  Furthermore, we provide analytic results for the
NLO QCD corrections, again expanded up to $1/m_t^8$. These expressions will be
of interest for cross checks of numerical results and for the construction of
approximation formulae involving expansions in different limits.

Calculations in the electroweak sector of the Standard Model are in general
much more complicated than in the strong sector since many different mass scales are
involved.  For the case of QCD corrections it has been
shown (see, e.g., Refs.~\cite{Davies:2019dfy,Bellafronte:2022jmo,Davies:2023vmj}) that
precise approximations
can be obtained by combining expansions performed in different regions
of the phase space.
This motivates developing these techniques beyond QCD to the electroweak
sector of the Standard Model.
In this work we take a first step in this direction by considering
the region in which the top quark mass is larger than all other kinematic
invariants.
While the radius of convergence of such an expansion is limited only
to small values of the centre-of-mass energy, the results will serve as
benchmarks for cross checks of other expansions or for numerical results.

This paper is organized as follows: in the next section we
define the form factors which describe the two processes considered,
and the technical details needed for our calculation are presented in
Section~\ref{sec::technical}. In particular, we describe the asymptotic
expansion and our renormalization procedure.  Section~\ref{sec::results}
contains our results for Higgs boson pair production and
Section~\ref{sec::results_gggh} is dedicated to the electroweak corrections to
$gg\to gH$. The QCD corrections to $gg\to gH$ are discussed in
Section~\ref{sec::results_gggh_QCD}.
In all cases we study the influence of the higher-order $1/m_t$ terms
on the form factors and provide our complete analytic expressions in the
ancillary files of this paper \cite{progdata}.
A brief summary of our findings is provided in
Section~\ref{sec::concl}.

%- }}}
%- {{{ Form factors for $gg\to HH$ and $gg\to gH$:

\section{Form factors for $gg\to HH$ and $gg\to gH$}

\subsection{$gg\to HH$}
The amplitude for the process
\begin{eqnarray}
  g(q_1)g(q_2)\to H(q_3)H(q_4)
\end{eqnarray}
can be decomposed into two Lorentz structures $A_1^{\mu\nu}$
and $A_2^{\mu\nu}$ which we define as
\begin{eqnarray}
  A_1^{\mu\nu} &=& g^{\mu\nu} - {\frac{1}{q_{12}}q_1^\nu q_2^\mu
  }\,,\nonumber\\
  A_2^{\mu\nu} &=& g^{\mu\nu}
                   + \frac{1}{{p_T^2} q_{12}}\left(
                   q_{33}    q_1^\nu q_2^\mu
                   - 2q_{23} q_1^\nu q_3^\mu
                   - 2q_{13} q_3^\nu q_2^\mu
                   + 2q_{12} q_3^\mu q_3^\nu \right)\,.
\end{eqnarray}
Here $q_{ij} = q_i\cdot q_j$ with $q_1^2=q_2^2=0$ and $q_3^2=q_4^2=m_H^2$.
$p_T$ is the transverse momentum of the final-state Higgs bosons, given by
\begin{eqnarray}
  p_T^2 &=&\frac{u\,t-m_H^4}{s}\,,
            \label{eq::pT}
\end{eqnarray}
with the Mandelstam variables
\begin{eqnarray}
  s = (q_1+q_2)^2\,,\qquad t = (q_1+q_3)^2\,,\qquad u = (q_1+q_4)^2\,.
  \label{eq::Mandelstam}
\end{eqnarray}
Using these definitions we introduce the form factors $F_1$ and $F_2$ as
\begin{eqnarray}
  {\cal M}^{ab} &=& 
  \varepsilon_{1,\mu}\varepsilon_{2,\nu}
  {\cal M}^{\mu\nu,ab}
  \,\,=\,\,
  \varepsilon_{1,\mu}\varepsilon_{2,\nu}
  \delta^{ab} X_0^{\rm ggHH} s 
  \left( F_1 A_1^{\mu\nu} + F_2 A_2^{\mu\nu} \right)
  \,,
                    \label{eq::M}
\end{eqnarray}
where $a,b$ are adjoint colour indices,
$X_0^{\rm ggHH} = G_F\alpha_s(\mu) T_F / (2\sqrt{2}\pi)$, $T_F=1/2$, $G_F$
is Fermi's constant and $\alpha_s(\mu)$ is the strong coupling constant
evaluated at the renormalization scale $\mu$.  We decompose the functions
$F_1$ and $F_2$ introduced in Eq.~(\ref{eq::M}) into ``triangle'' and ``box''
form factors. $F_1$ has contributions with zero, one and two $s$-channel
Higgs boson
propagators whereas $F_2$ only has box contributions. Thus we write
\begin{eqnarray}
  F_1 &=& \frac{3 m_H^2}{s-m_H^2}\left(
          F_{\rm tri}
          + \frac{m_H^2}{s-m_H^2} \tilde{F}_{\rm tri} 
          \right)
          + F_{\rm box1}
          \,, \nonumber\\
  F_2 &=& F_{\rm box2}\,.
                \label{eq::F_12}
\end{eqnarray}
In order to obtain this decomposition it is important to re-write the factors
of $s$ which occur in the numerators during the calculation using
$s/(s-m_H^2) = 1 + m_H^2/(s-m_H^2)$.  Note that at two loops $F_{\rm tri}$ is
not the same as the form factor for single Higgs boson production (as is
the case for QCD corrections), since loop corrections to the $HHH$ vertex also
enter here.

We define the perturbative expansion of the form factors as
\begin{eqnarray}
  F &=& F^{(0)} + \frac{\alpha_s(\mu)}{\pi} F^{(1,0)} + \frac{\alpha}{\pi} F^{(0,1)} 
        + \cdots
  \,,
  \label{eq::F}
\end{eqnarray}
where $\alpha$ is the fine structure constant and 
the ellipses indicate higher-order QCD and electroweak corrections.

In Section \ref{sec::results} we discuss the results for the squared
  matrix element constructed from the form factors $F_{\rm tri}$,
$\tilde{F}_{\rm tri}$,
$F_{\rm box1}$ and $F_{\rm box2}$.  Analytic results for the leading-order
form factors ($F_{\rm tri}^{(0)}$, $F_{\rm box1}^{(0)}$ and
$F_{\rm box2}^{(0)}$) are available from~\cite{Glover:1987nx,Plehn:1996wb}.
Two-loop corrections to $F_{\rm box1}^{(0,1)}$ and $F_{\rm box2}^{(0,1)}$
originating from the exchange of a virtual Higgs boson have been computed in
Ref.~\cite{Davies:2022ram} in the high-energy limit.

\begin{figure}[t]
  \centering
  \includegraphics[width=\textwidth]{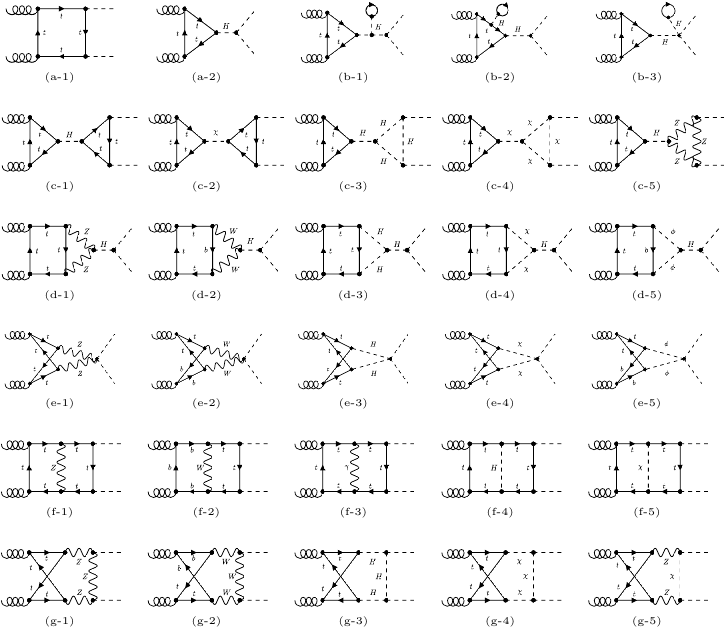}
  \caption{One- and two-loop Feynman diagrams contributing to $gg\to HH$.
  Dashed, solid, wavy and curly lines correspond to scalar particles,
  fermions, electroweak gauge bosons and gluons, respectively.}
  \label{fig::diags}
\end{figure}

In Fig.~\ref{fig::diags} we show sample one- and two-loop diagrams
contributing to $gg\to HH$. At two-loop order we have:
\begin{itemize}\setlength\itemsep{-.1em}
\item one-particle irreducible box and triangle diagrams,
\item one-particle reducible diagrams with a one-loop correction to the $HHH$
	vertex or a one-loop self-energy correction to the Higgs propagator of a one-loop $gg \to H \to HH$ diagram,
\item one-loop tadpole corrections to one-loop diagrams.
\end{itemize}

At two-loop order there are also contributions without top quarks which are
not suppressed by small Yukawa couplings.  In these contributions the
gluons couple to light quarks and the connection to the final-state Higgs
bosons is mediated via $Z$ bosons.  An example is given by diagram 
(g-1) in Fig.~\ref{fig::diags} if a light quark runs in
the fermion loop.  In our expansion these contributions formally contribute to
the $m_t^0$ term, however in this work we do not compute such diagrams;
they can be computed following the approach of Ref.~\cite{Bonetti:2020hqh}.

\begin{figure}[t]
  \centering
  \includegraphics[width=\textwidth]{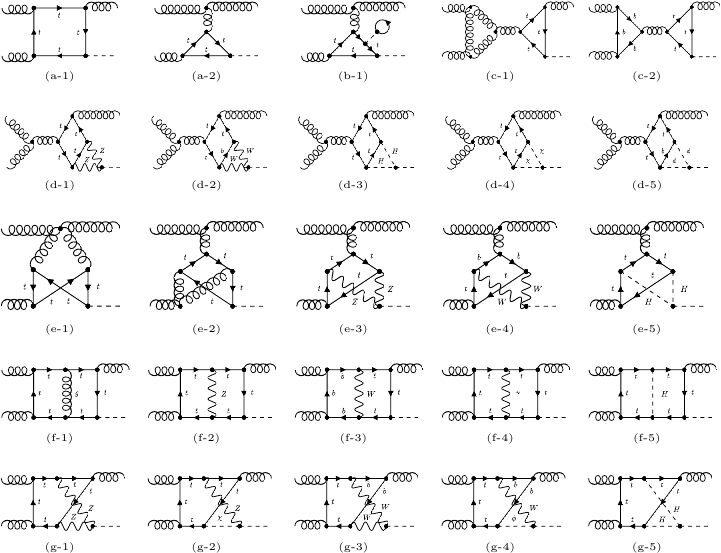}
  \caption{One- and two-loop Feynman diagrams contributing to $gg\to gH$.
    Dashed, solid, wavy and curly line correspond to scalar particles,
    fermions, electroweak gauge bosons and gluons, respectively.  
    Diagrams are also shown which contribute to the NLO QCD corrections.}
  \label{fig::diags_gggh}
\end{figure}

\subsection{$gg\to gH$}
The amplitude for the process
\begin{eqnarray}
  g(q_1)g(q_2)\to g(q_3)H(q_4)
\end{eqnarray}
can be decomposed into four physical Lorentz structures~\cite{Melnikov:2016qoc}
\footnote{We note that $A_4^{\mu\nu\rho}$ differs from Ref.~\cite{Melnikov:2016qoc} by the factor
of $1/s$, which we introduce such that all four form factors are dimensionless.}
\begin{eqnarray}
  A_1^{\mu\nu\rho} &= g^{\mu\nu} q_2^\rho \,, \qquad  A_2^{\mu\nu\rho} &= g^{\mu\rho} q_1^\nu \,, \nonumber \\
  A_3^{\mu\nu\rho} &= g^{\nu\rho} q_3^\mu \,, \qquad  
                     A_4^{\mu\nu\rho} &= \frac{1}{s} q_3^\mu q_1^\nu q_2^\rho \,.
\end{eqnarray}
The corresponding four form factors $F_1, \dots, F_4$ are defined through
\begin{eqnarray}
  {\cal M}^{abc} &=& f^{abc} X_0^{\rm gggH} 
                     \varepsilon_{1,\mu}\varepsilon_{2,\nu} \varepsilon_{3,\rho}
                     \sum_{i=1}^{4} F_i A_i^{\mu\nu\rho}
  \,,
  \label{eq::M2}
\end{eqnarray}
where $c$ is the adjoint colour index of the final-state gluon, $X_0^{\rm gggH}$ is given by
\begin{eqnarray}
   X_0^{\rm gggH} &=& 2^{1/4} \sqrt{4\pi\alpha_s(\mu) G_F} \, \frac{\alpha_s(\mu)}{4\pi}
\end{eqnarray}
and the perturbative expansions of the form factors are defined as in
\eqref{eq::F}. The Mandelstam variables are defined as in
Eq.~(\ref{eq::Mandelstam}); the only difference with respect to $gg \to HH$ is
that here $q_3^2=0$ and $p_T^2 = u\,t/s$.  Sample Feynman diagrams for
$gg\to gH$ are given in Fig.~\ref{fig::diags_gggh}.  The classification is
similar to $gg\to HH$, we again include all one-particle reducible and all
tadpole contributions.
Note that for the QCD corrections, we also include the one-loop self-energy corrections to the gluon propagators
and the one-loop vertex corrections to the triple-gluon vertex of the one-loop diagrams. The corrections to the quartic-gluon vertex do not appear at the two-loop order of this process.

%- }}}
%- {{{ Technical setup:

\section{\label{sec::technical}Technical setup}

%- {{{ Asymptotic expansion of the two-loop amplitudes:

\subsection{\label{sub::ae}Asymptotic expansion of the two-loop amplitudes}

For the generation of the $gg\to HH$ and $gg\to gH$ diagrams and the corresponding amplitudes
we use {\tt qgraf}~\cite{Nogueira:1991ex}. As input we use the Lagrangian file
of the full Standard Model shipped with {\tt tapir}~\cite{Gerlach:2022qnc},
which is derived from the Feynman rules of \texttt{UFO}~\cite{Darme:2023jdn}.
{\tt tapir} translates the {\tt qgraf} output
to {\tt FORM}~\cite{Ruijl:2017dtg} notation and generates further auxiliary
files which are useful for the manipulation of the amplitudes. The large-$m_t$
expansion is realized with the help of {\tt
  exp}~\cite{Harlander:1997zb,Seidensticker:1999bb} which generates the
corresponding subdiagrams and maps them to various integral
families.\footnote{See also Ref.~\cite{Smirnov:2023bxf} for a recent
  discussion of the expansion of integrals contributing to $H\to ggg$
  in the large-$m_t$ limit.}

\begin{figure}[t]
  \centering
  \includegraphics[width=\textwidth]{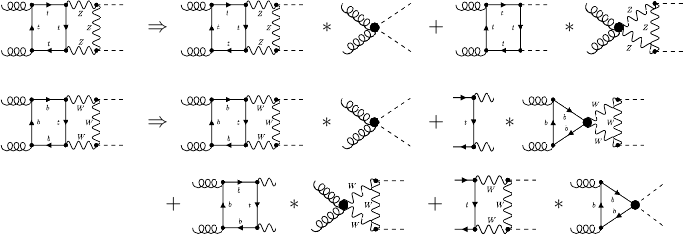}
  \caption{Asymptotic expansion of two sample Feynman diagrams.
    The subgraphs left of the stars have to be expanded in the small
    quantities, i.e., masses, external momenta or loop momenta
    of the co-subgraphs, which are  to the right of the stars.
  }
  \label{fig::ae}
\end{figure}

We apply the large-$m_t$ limit as
\begin{eqnarray}
  m_t^2 \gg s,t,m_W^2,m_Z^2,m_H^2
  \,,
  \label{eq::hierarchy}
\end{eqnarray}
where no additional hierarchy is assumed among the scales on the right-hand
side.  This leads to the following integral families:
\begin{itemize}\setlength\itemsep{-.1em}
\item one- and two-loop one-scale vacuum integrals,
\item one-loop massless triangle integrals where two external lines are massless,
\item massive vertex integrals where for one external leg we have $(q_1+q_2)^2=s$
  and for the other two legs we have $q_3^2=q_4^2=m_H^2$ (for $gg\to HH$)
  or $q_3^2=0$ and $q_4^2=m_H^2$ (for $gg\to gH$),
\item for the QCD corrections to $gg\to gH$ we also need massless one-loop box families
  with one external mass $q_4^2=m_H^2$; explicit analytic results can be
  found in Ref.~\cite{Ellis:2007qk}.
\end{itemize}
Our \texttt{FORM}-based setup automatically performs a reduction of arbitrary members
of each family to master integrals, which are well known in the literature (see,
e.g., Refs.~\cite{tHooft:1978jhc,Smirnov:2006vt}).
The tadpole integrals are computed by {\tt MATAD}~\cite{Steinhauser:2000ry} and the
remaining integral families are reduced by IBP reduction rules derived by \texttt{LiteRed}~\cite{Lee:2013mka}
which have been implemented in \texttt{FORM}.
Furthermore all of our reduction routines can deal with tensor integrals, avoiding
the need to construct additional projection operators.
In Fig.~\ref{fig::ae} we show how the various integral families appear due to the
asymptotic expansion in the large-$m_t$ limit.
In the Feynman gauge we have performed an expansion of the form factors up to order
$1/m_t^{10}$ ($1/m_t^{8}$) for $gg\to HH$ ($gg \to gH$).

In order to check our calculation, we also introduce general gauge parameters
$\xi_Z$, $\xi_W$ and $\xi_\gamma$ for the $Z$ and $W$ bosons and the
photon. From the technical point of view $\xi_\gamma$ does not introduce any
additional complexity since no new mass scale is introduced. It drops out
after summing all bare two-loop diagrams. This is not the case for $\xi_Z$ and
$\xi_W$ since they appear in combination with gauge boson masses in the gauge
boson and Goldstone propagators. Furthermore, $\xi_Z$ and $\xi_W$ only
drop out after renormalization. For this check we assume
\begin{eqnarray}
  m_t^2 \gg \xi_W m_W^2, \xi_Z m_Z^2 \gg s,t,m_W^2,m_Z^2, m_H^2 \,,
  \label{eq::hierarchy-xi}
\end{eqnarray}
and perform an expansion which includes terms up to order $1/m_t^4, 1/(\xi_W m_W^2)^2$,
$1/(\xi_Z m_Z^2)^2$, $1/(m_t^2 \xi_W m_W^2)$, $1/(m_t^2 \xi_Z m_Z^2)$ and
$1/(\xi_W m_W^2 \xi_Z m_Z^2)$.  To check the cancellation of $\xi_Z$ and
$\xi_W$ we have to consider the combination of the bare two-loop diagrams and
the counterterm contribution from the wave function of the external Higgs
boson (see also below), which also depends on $\xi_W$ and
$\xi_Z$.\footnote{Note that the counterterm contributions of the (physical)
  parameters are independent of the gauge parameters.}  It is a welcome and
non-trivial check of our calculation that up to this expansion depth,
$\xi_W$ and $\xi_Z$ drop out of the $gg\to HH$ and $gg\to gH$ amplitudes.

%- }}}
%- {{{ Renormalization:

\subsection{Renormalization}

In the following we concentrate on the electroweak sector; for the discussion
of the renormalization and the treatment of the infra-red divergences which
occur for the NLO QCD corrections to $gg\to gH$ we refer to
Section~\ref{sec::results_gggh_QCD}.

For the renormalization we follow the standard procedure as outlined, e.g., in
Refs.~\cite{Denner:1991kt,Denner:2019vbn}.  We express our one-loop amplitudes
for the form factors in terms of the parameters
\begin{eqnarray}
  e, m_W, m_Z, m_t, m_H,
\end{eqnarray}
where $e = \sqrt{4\pi\alpha}$, and introduce one-loop on-shell counterterms (see,
e.g. Eqs.~(143),~(153) and~(421) of Ref.~\cite{Denner:2019vbn}). Furthermore,
we have to renormalize the wave function of the external Higgs boson, which we
also perform in the on-shell scheme (see Eq.~(144) of
Ref.~\cite{Denner:2019vbn}).

We consistently include tadpole contributions in all parts of our calculation
(in the two-loop $gg\to HH$ and $gg\to gH$ amplitudes, and the gauge boson and fermion two-point functions
needed for the counterterms). This guarantees that the top quark mass
counterterm is gauge-parameter independent.  This prescription is equivalent
to the so-called {\it Fleischer–Jegerlehner tadpole
  scheme}~\cite{Fleischer:1980ub}.\footnote{For a recent detailed
  discussion on the various tadpole renormalization schemes
  we refer to Ref.~\cite{Dittmaier:2022maf}.}

For the numerical evaluation of the form factors we transform our results into
the so-called $G_\mu$ scheme where the Fermi constant $G_F$ and the gauge
boson masses  $m_Z$ and $m_W$ are the input
parameters, and the fine structure constant $\alpha$ and the weak mixing angle
$\theta_W$ are derived quantities.  (see,
e.g., Section~5.1.1 of Ref.~\cite{Denner:2019vbn}). In this scheme it is 
convenient to express the final result in terms of the variable
\begin{eqnarray}
  x_t &=& \frac{G_F m_t^2}{8\sqrt{2}\pi^2}\,.
\end{eqnarray}

Although we have computed the exact top quark mass dependence of all counterterm
contributions it is convenient to expand them in $1/m_t$ and combine the
individual terms with the expanded bare two-loop amplitude.
We do not expand the (finite) quantity $\Delta r$,
which performs the transformation from the $\alpha$ to the $G_\mu$ scheme,
in the large-$m_t$ limit but retain its exact dependence on $m_t$.

Note that the NLO electroweak corrections do not produce infra-red divergences.
Thus, already after renormalization we obtain the finite results for the form factors.
This is not the case for the NLO QCD corrections to $gg\to gH$; the infra-red subtraction
necessary to produce a finite result is discussed in Section \ref{sec::results_gggh_QCD}.
Let us also mention that our NLO electroweak form factors do not have an explicit
dependence on the renormalization scale since all parameters are
renormalized in the on-shell scheme.

%- }}}

%- }}}
%- {{{ Results ggHH:

\section{\label{sec::results}Results for $gg\to HH$}

%- {{{ Analytic results:

\subsection{\label{sub::ana}Analytic results}

It is instructive to begin by discussing the leading contributions in the
large-$m_t$ expansion, of order $m_t^4$ and $m_t^2$, which are present in
$F^{(0,1)}_{\rm tri}$ and $F^{(0,1)}_{\rm box1}$.
Our results for the two-loop form factors read
\begin{eqnarray}
  \frac{\alpha}{\pi} F^{(0,1)}_{\rm tri} 
  &=& \frac{4}{3} \times x_t 
      \left(
      \frac{136}{15} - \frac{16m_t^2}{m_H^2} 
      \right)
      + {\cal O}\left(m_t^0\right) \,,\nonumber\\
  \frac{\alpha}{\pi} F^{(0,1)}_{\rm box1} 
  &=& - \frac{4}{3} \times \frac{4 x_t}{5}
      + {\cal O}\left(m_t^0\right) \,.
      \label{eq::Ftri_box_mt4}
\end{eqnarray}
For reference, we also provide
the large-$m_t$ limit of the leading-order form factors which are given by
\begin{eqnarray}
  F^{(0)}_{\rm tri}  &=& \frac{4}{3}  + {\cal O}\left(1/m_t^2\right)\,,\nonumber\\
  F^{(0)}_{\rm box1} &=& -\frac{4}{3} + {\cal O}\left(1/m_t^2\right)\,.
\end{eqnarray}

Results for $F_{\rm tri}$ and $F_{\rm box1}$ have also been presented
in Ref.~\cite{Muhlleitner:2022ijf}, in which leading $m_t^2$
corrections to the $ggH$ and $ggHH$ vertices at two-loop order are
taken into account using an effective-theory approach, while
one-particle reducible diagrams have been computed with full $m_t$
dependence.  Furthermore, all one- and two-particle reducible diagrams
involving Yukawa couplings have been considered. 
After extracting the $m_t^4$ and $m_t^2$ terms we find agreement with our
results. To make this comparison it is important to consider sub-leading terms
in the expansion of the LO form factors which are factored out in
Ref.~\cite{Muhlleitner:2022ijf} and contain exact $m_t$ dependence.

Using the asymptotic expansion described in Section~\ref{sub::ae} we have
obtained expansion terms up to order $1/m_t^{10}$. Up to order $1/m_t^4$ we
have performed the calculation for general gauge parameters and we have
verified that they drop out from the renormalized results.  The higher-order
$1/m_t$ terms have been computed only in the Feynman gauge. The analytic expressions
for the form factors can be obtained from~\cite{progdata}.

In our analytic expressions we observe poles of the form $1/(s-4m_H^2)^k$ where
$k>0$ is larger for the higher-order $1/m_t$ terms. The origin of these terms
are massive one-loop triangle (co-)subgraphs,
such as the one on the first row of
Fig.~\ref{fig::ae} with external squared momenta $s$, $m_H^2$ and $m_H^2$. The
expansion of the subgraph leads to numerators in the triangle diagram and the
$1/(s-4m_H^2)$ terms result from the subsequent reduction to master integrals.
We note that the poles are spurious; for each $1/m_t$ term
the limit $s\to 4 m_H^2$ exists.

We also point out that the $m_t^0$ term presented here is not complete,
since it should also receive contributions from
diagrams without top quarks, for e.g., the first diagram in
Fig.~\ref{fig::ae} where the top quarks are replaced by light quarks.  We do
not compute such diagrams in this paper. They can be computed following the
approach of, e.g., Ref.~\cite{Bonetti:2020hqh} where similar contributions to
$gg\to gH$ have been considered, or with the help of expansions as proposed,
e.g. in Ref.~\cite{Davies:2022ram}.

%- }}}
%- {{{ Numeric results:

\subsection{\label{sub::num}Numeric results}

For the numerical evaluation of our form factors we adopt the $G_\mu$
scheme and use the following input values
\begin{eqnarray}
  m_t = 172~\mbox{GeV}\,, &&  m_H = 125~\mbox{GeV}\,, \nonumber\\
  m_W = 80~\mbox{GeV}\,, && m_Z = 91~\mbox{GeV}\,.
                            \label{eq::num}
\end{eqnarray}
Furthermore, we express the form factors in terms of $s$ and $p_T$
and introduce the parameter
\begin{eqnarray}
  \rho_{p_T} = \frac{p_T}{\sqrt{s}}
  \,.
\end{eqnarray}
In the following we choose $\rho_{p_T}=0.1$ and
discuss results for the squared matrix element 
\begin{eqnarray}
  {\cal U}_{\rm ggHH} \equiv 
       \frac{1}{8^2} \sum_{\rm col} 
       \frac{1}{2^2} \sum_{\rm pol} |{\cal M}^{ab}|^2 
  = \frac{1}{16}
  \left( X_0^{\rm ggHH} s \right)^2 \left( |F_1|^2 + |F_2|^2 \right)
  = \frac{1}{16} 
  \left( X_0^{\rm ggHH} s \right)^2 \tilde{ {\cal U} }_{\rm ggHH} 
  \,.
\end{eqnarray}
For the numerical evaluation of the massive two- and three-point functions
we use the program \texttt{Package-X}~\cite{Patel:2016fam}.

\begin{figure}[t]
  \centering
  \begin{tabular}{cc}
    \includegraphics[width=.45\textwidth]{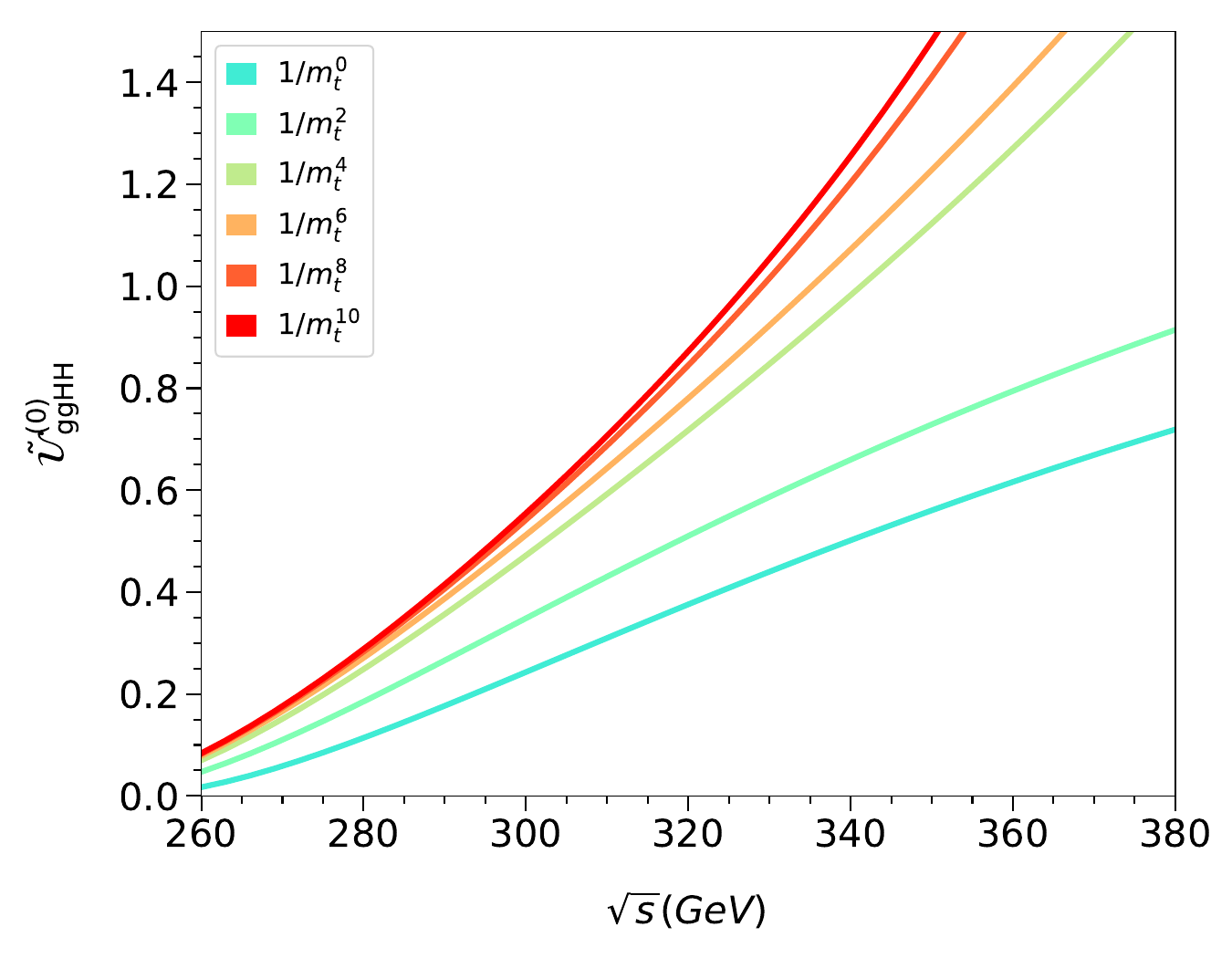} &
    \includegraphics[width=.45\textwidth]{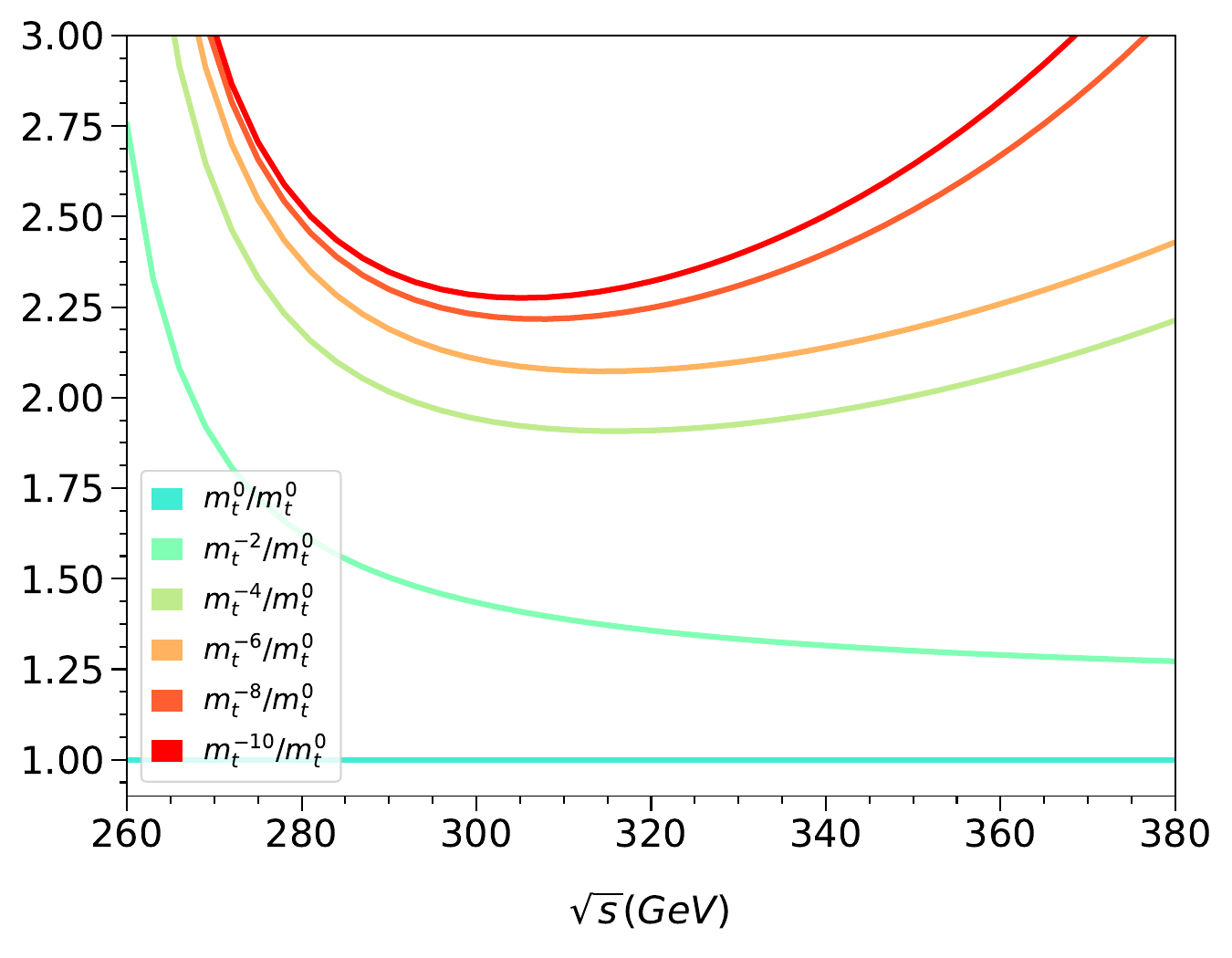}
  \end{tabular}
  \caption{\label{fig::gghh_LO}
      $\tilde{{\cal U}}_{\rm ggHH}^{(0)}$ plotted as a function of
      $\sqrt{s}$. Results are shown up to order $1/m_t^{10}$.
        The panel on the right shows the result normalized to the
        $m_t^0$ expansion term.}
\end{figure}

\begin{figure}[t]
  \centering
  \begin{tabular}{cc}
    \includegraphics[width=.45\textwidth]{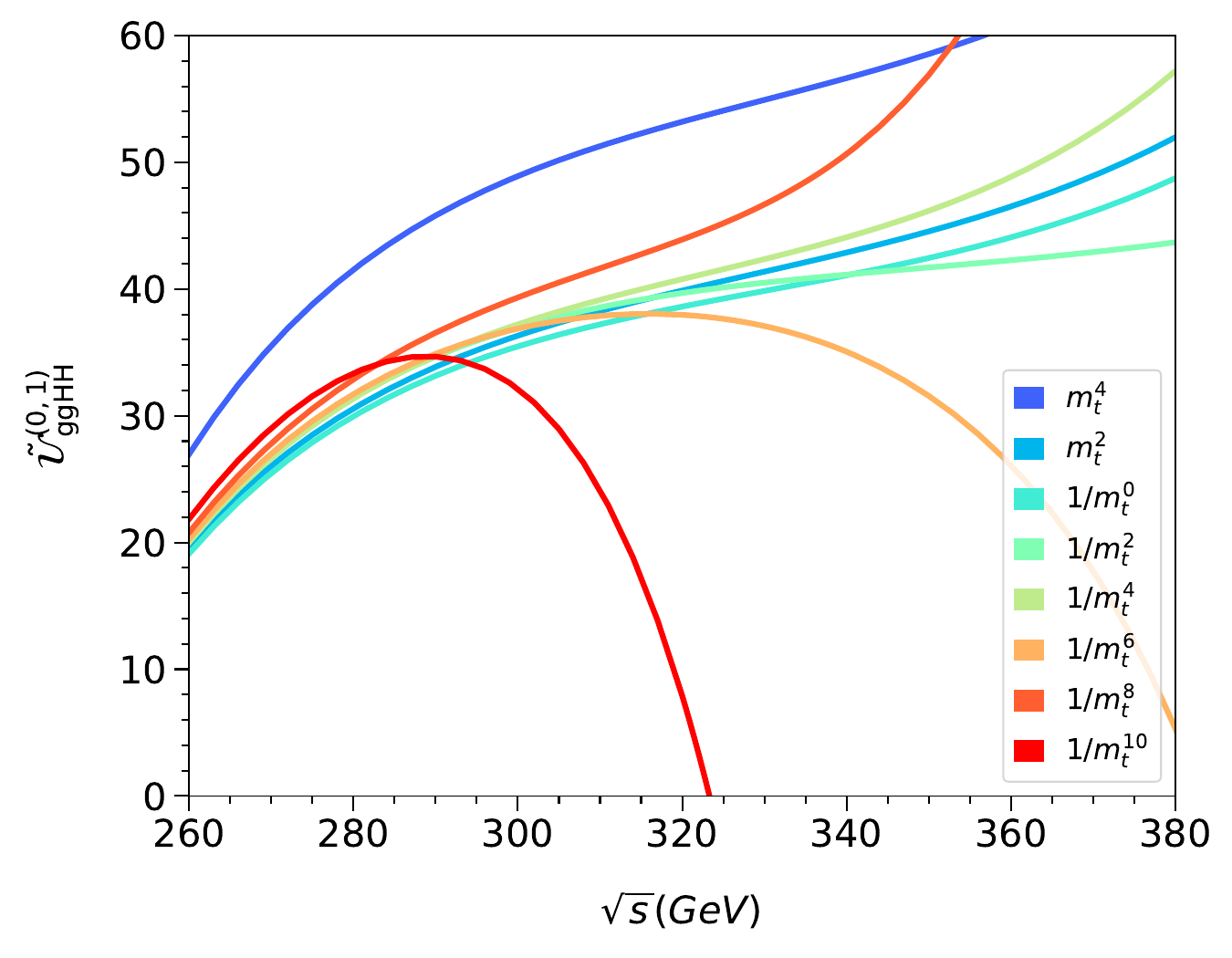} &
    \includegraphics[width=.45\textwidth]{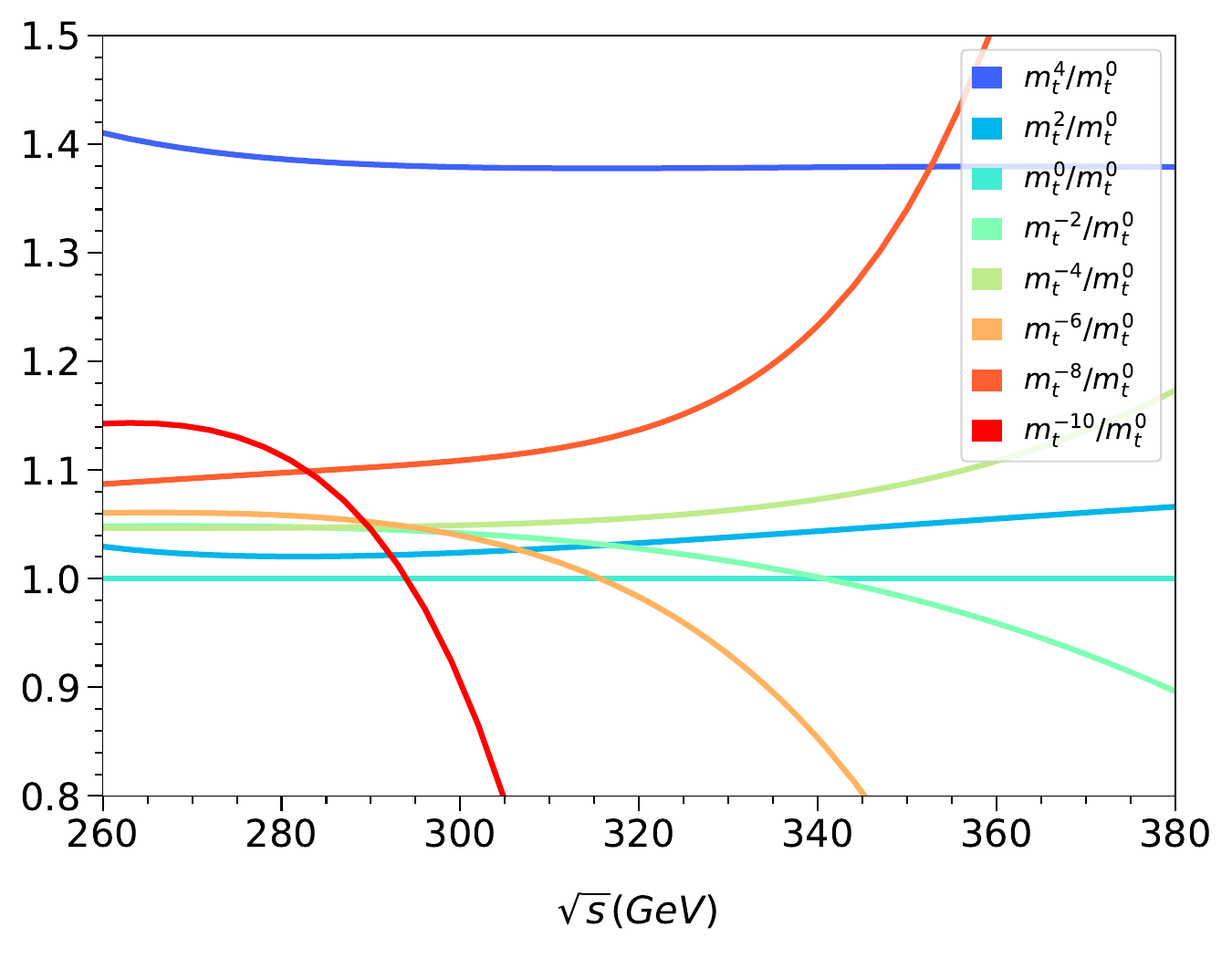}
  \end{tabular}
  \caption{\label{fig::gghh_NLO}$\tilde{{\cal U}}_{\rm ggHH}^{(0,1)}$ as a
    function of $\sqrt{s}$. The panel on the right shows the result
      normalized to the $m_t^0$ expansion term.}
\end{figure}

\begin{figure}[t]
  \centering
  \begin{tabular}{cc}
    \includegraphics[width=.45\textwidth]{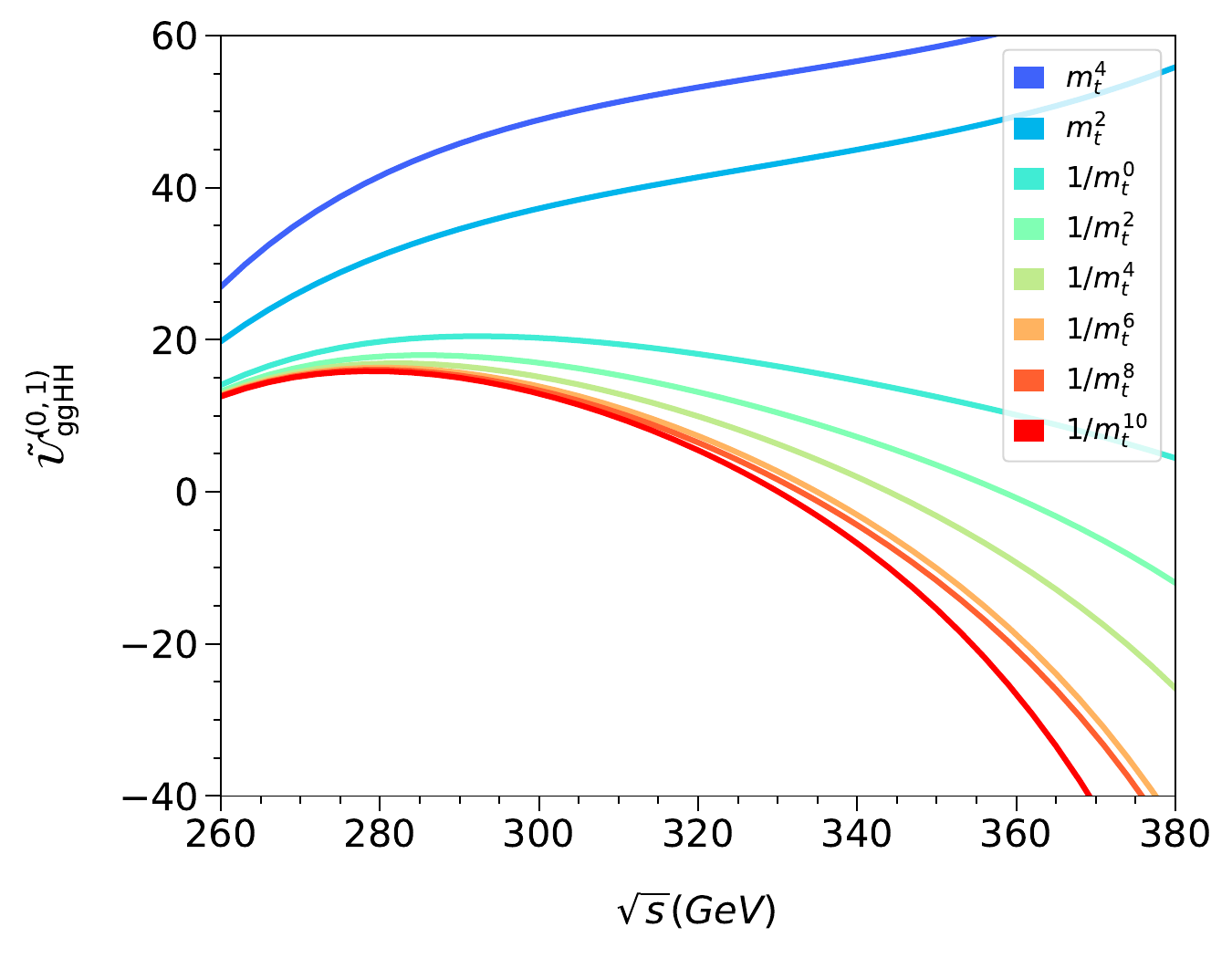} &
    \includegraphics[width=.45\textwidth]{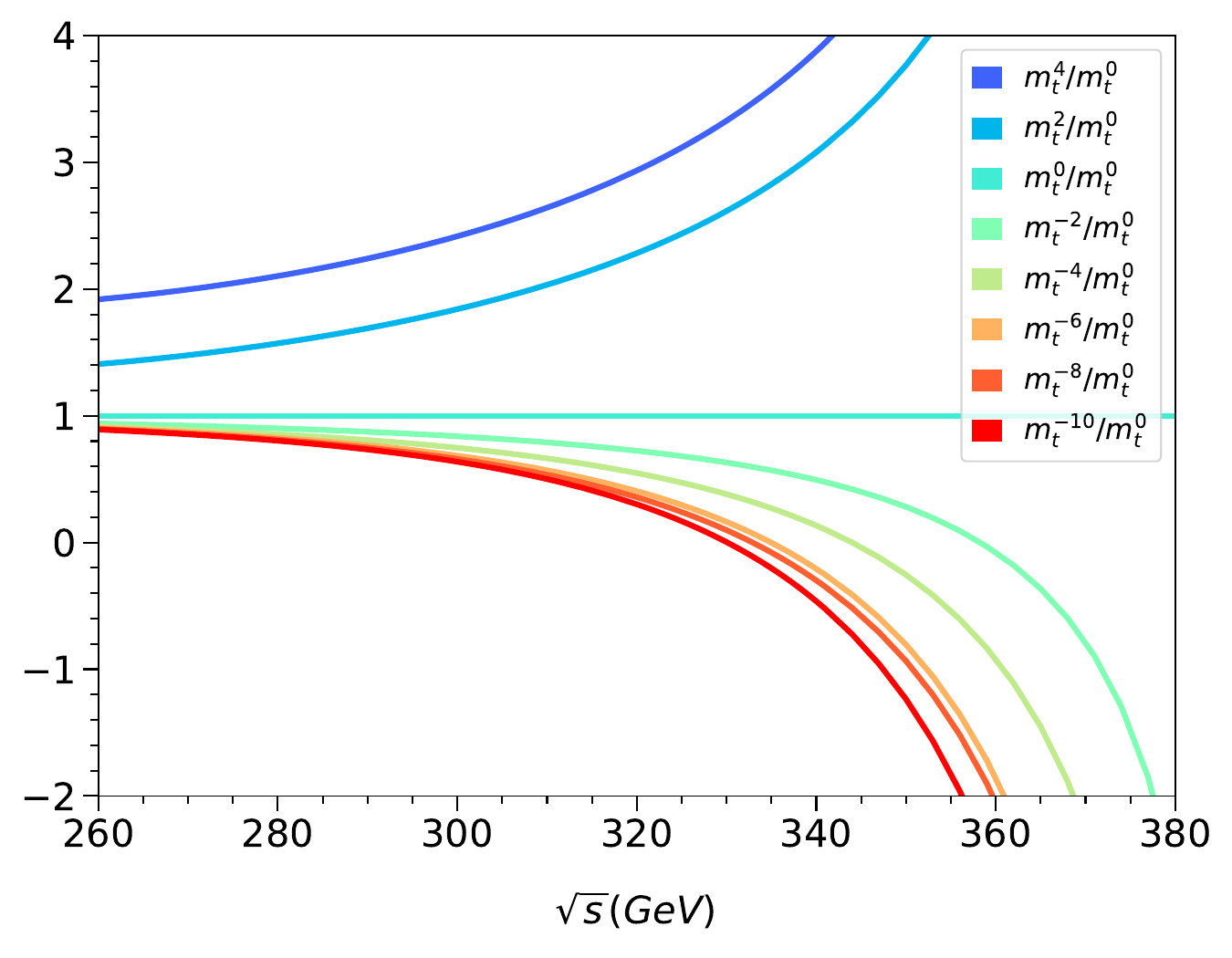}
  \end{tabular}
  \caption{\label{fig::gghh_NLO_no_tbW}$\tilde{{\cal U}}_{\rm ggHH}^{(0,1)}$
    without contributions involving a cut at $\sqrt{s} = m_t+m_W$, see text
    for details. The panel on the right shows the result normalized to the
      $m_t^0$ expansion term.}
\end{figure}

For reference, in Fig.~\ref{fig::gghh_LO} we show the LO contribution to
$\tilde{{\cal U}}_{\rm ggHH}$ as a function of $\sqrt{s}$.
Below the top quark threshold the expansion converges well, however
it converges more slowly as $\sqrt{s}$ gets closer to $2m_t$.

In Fig.~\ref{fig::gghh_NLO} we show the NLO quantity
$\tilde{{\cal U}}_{\rm ggHH}^{(0,1)}$ as a function of $\sqrt{s}$.  The curves
include increasing expansion depths starting from the leading term
proportional to $m_t^4$ (which originates from $F_{\rm tri}^{(0,1)}$) up to
$1/m_t^{10}$.  For the $\sqrt{s}$ axis we choose values from the Higgs pair
production threshold at $2 m_H = 250$~GeV up to $\sqrt{s}=380$~GeV. Note
that convergence of the expansion is not expected beyond the top quark pair
production threshold at $2 m_t = 344$~GeV.  Below this value we observe, at
first sight, a reasonable convergence. Below $\sqrt{s}\approx 300$~GeV a
significant shift is obtained from the constant contribution proportional to
$m_t^0$ and higher order $1/m_t$ terms are small up to $1/m_t^8$. However, the
$1/m_t^{10}$ contribution again provides a sizeable shift, which is clearly
visible on the right panel which shows the ratio with respect to the $m_t^0$
contribution.

This behaviour is due to diagrams with a closed quark loop
which contains both top and bottom quarks, see, e.g., the second diagram in
Fig.~\ref{fig::ae}. Such diagrams contain cuts through a top quark and $W$ boson
and thus the large-$m_t$ expansion is expected to break down above
$\sqrt{s} = m_t+m_W \approx 250$~GeV. Diagrams with such a cut contribute to
both $F_1$ and $F_2$.
To demonstrate this,
in Fig.~\ref{fig::gghh_NLO_no_tbW} we show the results for
$\tilde{{\cal U}}_{\rm ggHH}^{(0,1)}$ where we set all diagrams containing a bottom quark to
zero in the finite parts.\footnote{The $1/\epsilon$ poles parts are required in order to
  obtain finite expressions after renormalization.}
We indeed observe that after removing these contributions the large-$m_t$ expansion
converges as expected up to the threshold at $\sqrt{s}=2 m_t$.
We note that the two-loop diagrams have further cuts where no top quark
is involved at $\sqrt{s}=2m_W, 2 m_Z, 2 m_H$. In our approach all of
these are taken into account exactly, so they do not affect the convergence of
the large-$m_t$ expansion.

In view of the above discussion the validity of the leading $m_t$
terms (see Section~\ref{sub::ana} and Ref.~\cite{Muhlleitner:2022ijf}), and
indeed of the deeper large-$m_t$ expansion, for a description of the
electroweak corrections to $gg\to HH$ is questionable.
More insight will be provided in a future publication
which considers the small-$t$ expansion of these diagrams in the style of
Ref.~\cite{Davies:2023vmj}.

%- }}}

%- }}}
%- {{{ Results gggH - ew:

\section{\label{sec::results_gggh}Results for $gg\to gH$: electroweak corrections}

In this section we consider the electroweak corrections to $gg\to gH$.
The QCD corrections are presented in Section~\ref{sec::results_gggh_QCD}.
For the input values for numerical evaluation we adopt the values given
in Eq.~(\ref{eq::num}).

In order to study the convergence of the expansion in $1/m_t$ we 
consider the squared matrix element since the individual form factors show a
divergent behaviour for $s\to m_H^2$ which is due to contributions where a
gluon is present in the $t$ or $u$ channel. In principle one could further
decompose the form factors to make this dependence explicit, however, we
prefer to consider
\begin{eqnarray}
  {\cal U}_{\rm gggH} &\equiv& 
   \frac{1}{8^2} \sum_{\rm col} 
   \frac{1}{2^2} \sum_{\rm pol} |{\cal M}^{abc}|^2
                               \nonumber\\
  &=& \frac{3}{32}  \left( X_0^{\rm gggH} \right)^2\Bigg\{
      s \left[\frac{2 F_1 F_1^\star u}{t}+\frac{2 F_2 F_2^\star t}{u} +F_2 F_1^\star+F_1 F_2^\star\right]
      \nonumber\\&&\mbox{}
      +\Big[F_4 \left(F_3^\star+F_4^\star\right) +F_3 \left(2 F_3^\star+F_4^\star\right)\Big] \frac{t\,u}{s}
      \vphantom{\Bigg\}}\nonumber\\&&\mbox{}
      +\Big[\left(F_3+F_4\right) F_2^\star+F_2 \left(F_3^\star+F_4^\star\right)\Big] t
      \vphantom{\Bigg\}}\nonumber\\&&\mbox{}
      +\Big[\left(F_3+F_4\right) F_1^\star+F_1 \left(F_3^\star+F_4^\star\right)\Big] u 
      \Bigg\}
  \nonumber\\
  &=&  \frac{3}{32} 
      { \left( X_0^{\rm gggH} \right)^2 s \,\, 
      \tilde{ {\cal U} }_{\rm gggH} }\,,
\end{eqnarray}
where $F_i^\star$ denotes the complex-conjugate form factors.
After inserting the perturbative expansion from Eq.~(\ref{eq::F})
we obtain the LO and NLO contributions to ${\cal U}_{\rm gggH}$,
which converge for $s\to m_H^2$.

\begin{figure}[t]
  \centering
  \begin{tabular}{cc}
    \includegraphics[width=.45\textwidth]{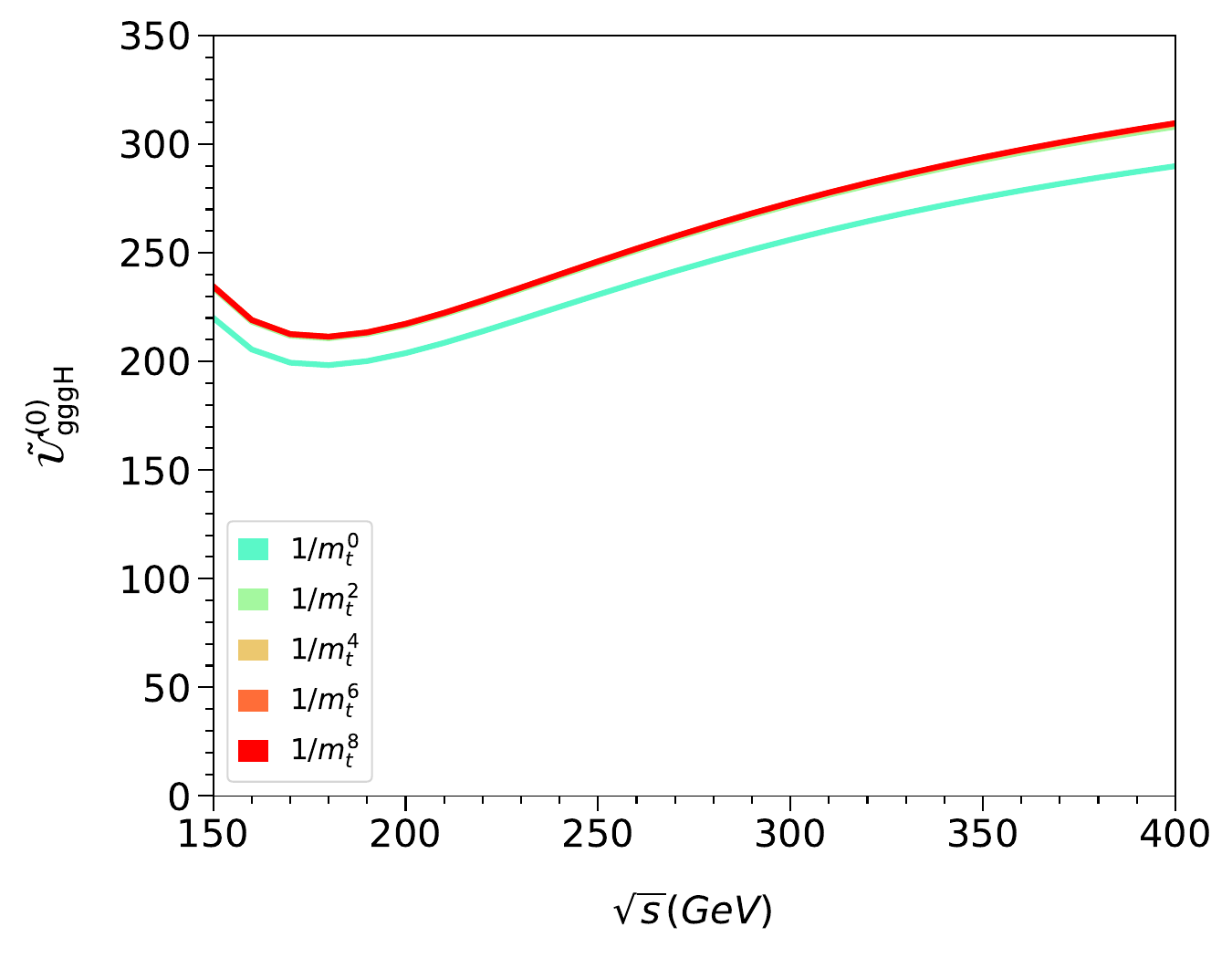} &
    \includegraphics[width=.45\textwidth]{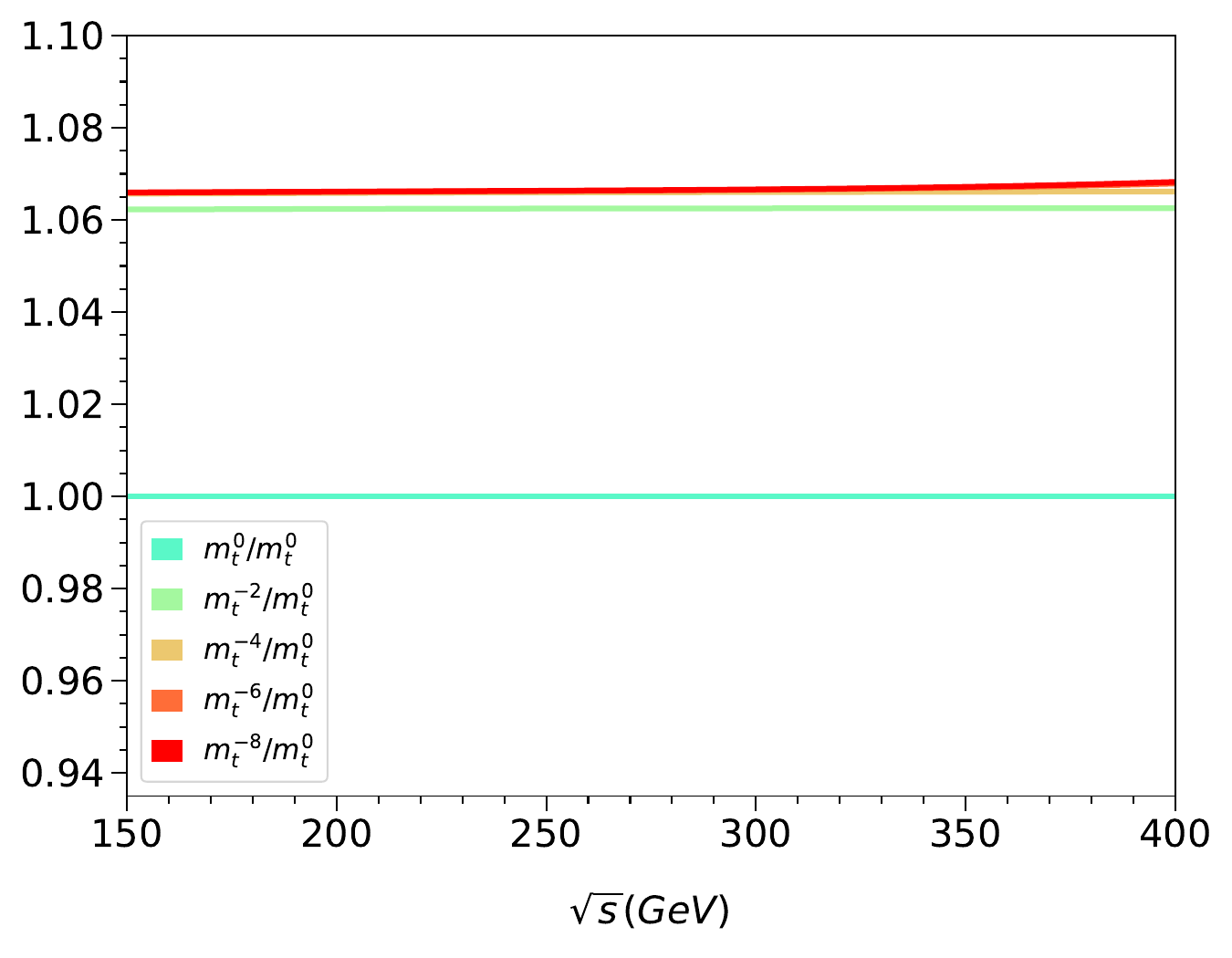}
  \end{tabular}
  \caption{\label{fig::gghg_1l}Left: $\tilde{{\cal U}}_{\rm gggH}^{(0)}$ as a function
    of $\sqrt{s}$. Right: Ratio with respect to the 
    $m_t^0$ expansion term.  The various colours correspond to the
    inclusion of different expansion terms.}
\end{figure}

We start with the discussion of the LO corrections. In Fig.~\ref{fig::gghg_1l}
we show $\tilde{{\cal U}}_{\rm gggH}^{(0)}$, for $\rho_{p_T}=0.1$, as a function
of $\sqrt{s}$. The right panel shows the ratio with respect to the leading
expansion term. We observe very good convergence below $\sqrt{s}=2m_t$ and can
safely assume that we reproduce the exact result every time two successive
expansion terms overlap.  In fact, below $\sqrt{s}\approx 250$~GeV only the
first three terms lead to visible shifts and below $\sqrt{s}\approx 300$~GeV
the curve which includes $1/m_t^8$ terms (which is the order we have available
at two loops) provides a good approximation. The inclusion of $1/m_t^{14}$
terms extends the convergence region even further. The one-loop form factors
enter the construction of $\tilde{{\cal U}}_{\rm gggH}^{(0,1)}$; due to their
excellent convergence it is safe to use the expansion, including terms to
$1/m_t^{14}$, and avoid implementing the exact, analytic leading-order
expression.

\begin{figure}[t]
  \centering
  \begin{tabular}{cc}
    \includegraphics[width=.45\textwidth]{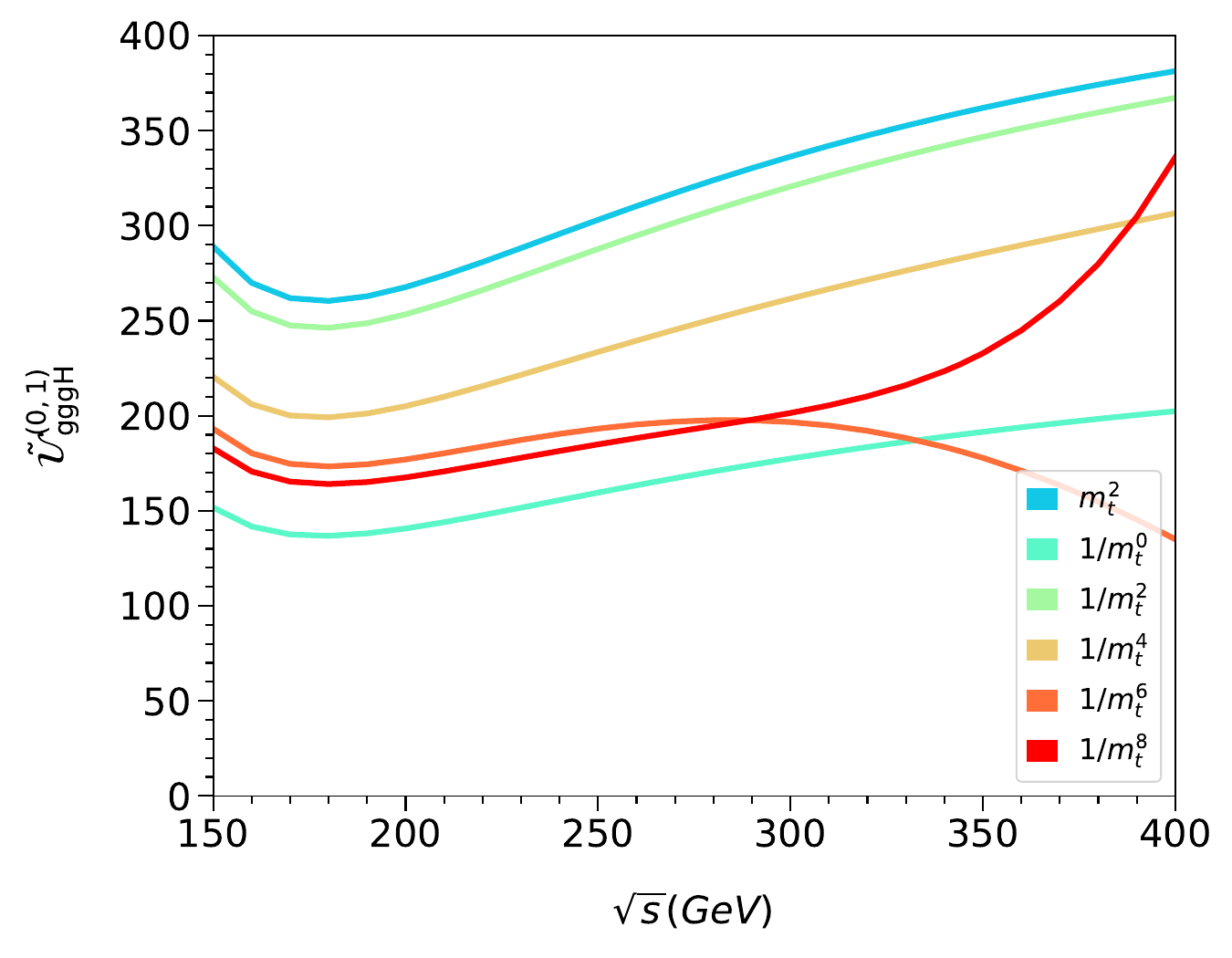} &
    \includegraphics[width=.45\textwidth]{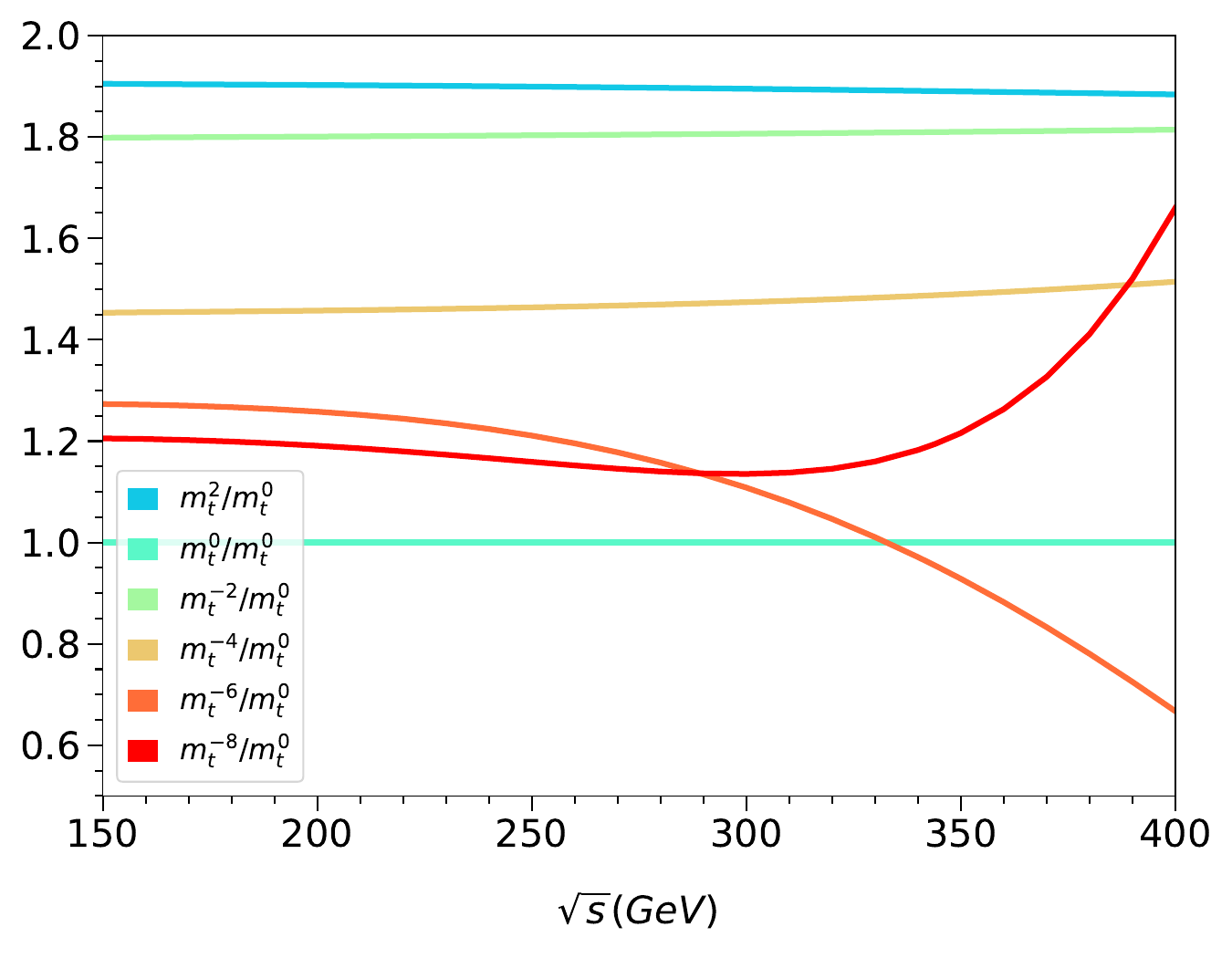}
  \end{tabular}
  \caption{\label{fig::gghg_2l} $\tilde{{\cal U}}_{\rm gggH}^{(0,1)}$ as a function of
    $\sqrt{s}$. The ratio with respect to the
    $m_t^0$ expansion term is shown in the right panel.}
\end{figure}

NLO results for $\tilde{{\cal U}}_{\rm gggH}$ in the $G_\mu$ scheme are shown
in Fig.~\ref{fig::gghg_2l}, again for $\rho_{p_T}=0.1$.  As expected, we
observe good convergence below the top quark threshold. In particular below
$\sqrt{s}\approx 300$~GeV the higher order $1/m_t$ terms become smaller and
smaller and the approximation which includes $1/m_t^8$ terms agrees well with
the $1/m_t^6$ approximation.  From the right panel we observe that the
$\{1/m_t^2,1/m_t^4,1/m_t^6\}$ terms lead an almost $s$-independent shift of
about $\{80\%,20\%,10\%\}$ and the $1/m_t^8$ term provides only a shift at the
few-percent level.

We have compared our one-loop form factors to Ref.~\cite{Gao:2023bll} and find
agreement up to $1/m_t^{14}$.  We also compare with the subset of
NLO contributions induced by the trilinear Higgs boson coupling considered in
Ref.~\cite{Gao:2023bll}, by extracting the corresponding pieces from our
bare two-loop form factors. We have compared up to $1/m_t^{2}$ and find
agreement.

Our result provides solid predictions for the energy range
$m_H\le \sqrt{s} \lesssim 300$~GeV and will thus serve as
an important cross check for future (analytic) calculations
in different kinematic limits or of numerical evaluations.

%- }}}
%- {{{ Results gggH - QCD:

\section{\label{sec::results_gggh_QCD}NLO QCD corrections to $gg\to gH$ in the
  large-$m_t$ limit}

A finite expression for the NLO virtual QCD corrections to $gg\to gH$
is obtained after introducing counterterms for the ultra-violet
poles and subtracting the infra-red divergences.
We first renormalize the strong coupling constant in
the $\overline{\rm MS}$ scheme with six active flavours. The top
quark mass and gluon wave functions are renormalized in the on-shell
scheme.\footnote{The transition from the on-shell to the $\overline{\rm MS}$
  quark mass is straightforward.} Afterwards we express the form factors in
terms of $\alpha_s^{(5)}(\mu)$, with five active flavours.
Finite form factors are then obtained via the subtraction ($i=1,2,3,4$)
\begin{eqnarray}
  F_{i,\mathrm{fin}}^{(1,0)} &=& F_{i,\mathrm{ren}}^{(1,0)} - \frac{1}{2} I_g^{(1)} F_{i}^{(0)}
                                 \label{eq::F_fin}
\end{eqnarray}
where $F_{i,\mathrm{ren}}^{(1,0)}$ are the ultra-violet renormalized form factors.
The quantity $I_g^{(1)}$ on the right-hand side of Eq.~(\ref{eq::F_fin})
is given by~\cite{Catani:1998bh}
\begin{eqnarray}
  I_g^{(1)} 
  &=&
      - 
      \frac{  e^{\epsilon \gamma_E}}{ {2} \Gamma (1-\epsilon)}
      \left(\frac{C_A}{\epsilon^2}+ \frac{2\beta_0}{\epsilon}\right)
      \left[\left(\frac{\mu^2}{-s-i\delta}\right)^{\epsilon} 
      + \left(\frac{\mu^2}{-t}\right)^{\epsilon} 
      + \left(\frac{\mu^2}{-u}\right)^{\epsilon}\right] \,, 
\end{eqnarray}
with $\beta_0 = 11C_A/12 - T_F\,n_l/3$, where $T_F=1/2$, $C_A = n_c$ and $n_l$ is
the number of massless quarks.

For illustration we present the one- and two-loop expressions for the form factors
$F_{1}^{(0)}$ and $F_{1,\mathrm{fin}}^{(1,0)}$ to the expansion order
$1/m_t^2$ and $m_t^0$, respectively. Deeper expansions can be found in
the supplementary material~\cite{progdata} of this paper. At one-loop order we have
\begin{eqnarray}
F_{1}^{(0)} &=&
\frac{(s+t) \left(m_h^2-t\right)}{3 s u}  \left[ 
-4
+ \frac{1}{m_t^2}\left(\frac{7 m_h^4 (s+t)-t m_h^2 (10 s+7 t)+3 s t (s+t)}{30 (s+t) \left(t-m_h^2\right)}\right)
\right],
\nonumber \\
\end{eqnarray}
and the two-loop expression is given by
\begin{eqnarray}
F_{1,\mathrm{fin}}^{(1,0)} &\!\!=\!\!&
\frac{(s+t) (m_h^2-t)}{3 s u} \bigg( -\frac{3}{2 n_c}
+ n_c \bigg\{2 \text{Li}_2\Big(1-\frac{s}{m_h^2}\Big)-2 \text{Li}_2\Big(\frac{t}{m_h^2}\Big)-2 \text{Li}_2\Big(\frac{u}{m_h^2}\Big) \nonumber \\
&&{}+\frac{m_h^2 (21 s+23 t)-23 t (s+t)}{6 (s+t) (t-m_h^2)}+\log ^2\Big(\frac{s}{m_h^2}\Big)+\log ^2\Big(\!-\frac{t}{m_h^2}\Big) \nonumber \\
&&{}+2 i \pi  \log \Big(\!-\frac{t}{m_h^2}\Big)-\bigg[2 \log \Big(\!-\frac{t}{m_h^2}\Big)+2 i \pi \bigg] \log \Big(1-\frac{t}{m_h^2}\Big)+\log ^2\Big(\!-\frac{u}{m_h^2}\Big) \nonumber \\
&&{}+2 i \pi  \log \Big(\!-\frac{u}{m_h^2}\Big)-\bigg[2 \log \Big(\!-\frac{u}{m_h^2}\Big)+2 i \pi \bigg] \log \Big(1-\frac{u}{m_h^2}\Big)-\log ^2\Big(\frac{\mu ^2}{s}\Big) \nonumber \\
&&{}+\log \Big(\frac{\mu ^2}{s}\Big) \bigg[\log \Big(\!-\frac{\mu ^2}{t}\Big)+\log \Big(\!-\frac{\mu ^2}{u}\Big)-\frac{11}{6}-2 i \pi \bigg] \nonumber \\
&&{}-\log ^2\Big(\!-\frac{\mu ^2}{t}\Big)+\log \Big(\!-\frac{\mu ^2}{t}\Big) \bigg[\log \Big(\!-\frac{\mu ^2}{u}\Big)-\frac{11}{6}+i \pi \bigg] \nonumber \\
&&{}-\log ^2\Big(\!-\frac{\mu ^2}{u}\Big)-\bigg[\frac{11}{6}-i \pi \bigg] \log \Big(\!-\frac{\mu ^2}{u}\Big)-\frac{5 \pi ^2}{6}-\frac{11 i \pi }{6}\bigg\}  +\log \Big(\frac{\mu ^2}{m_t^2}\Big) \nonumber \\
%%%%%
&&{}+n_l \bigg\{ \frac{t u}{3 (s+t) (m_h^2-t)}+\frac{1}{3} \log \Big(\frac{\mu ^2}{s}\Big)+\frac{1}{3} \log \Big(\!-\frac{\mu ^2}{t}\Big)+\frac{1}{3} \log \Big(\!-\frac{\mu ^2}{u}\Big)+\frac{i \pi }{3}\bigg\}\!\bigg)\,,
\nonumber \\
\end{eqnarray}
where $n_c=3$ and $\text{Li}_2$ is the dilogarithm.

\begin{figure}[t]
  \centering
  \begin{tabular}{cc}
    \includegraphics[width=.45\textwidth]{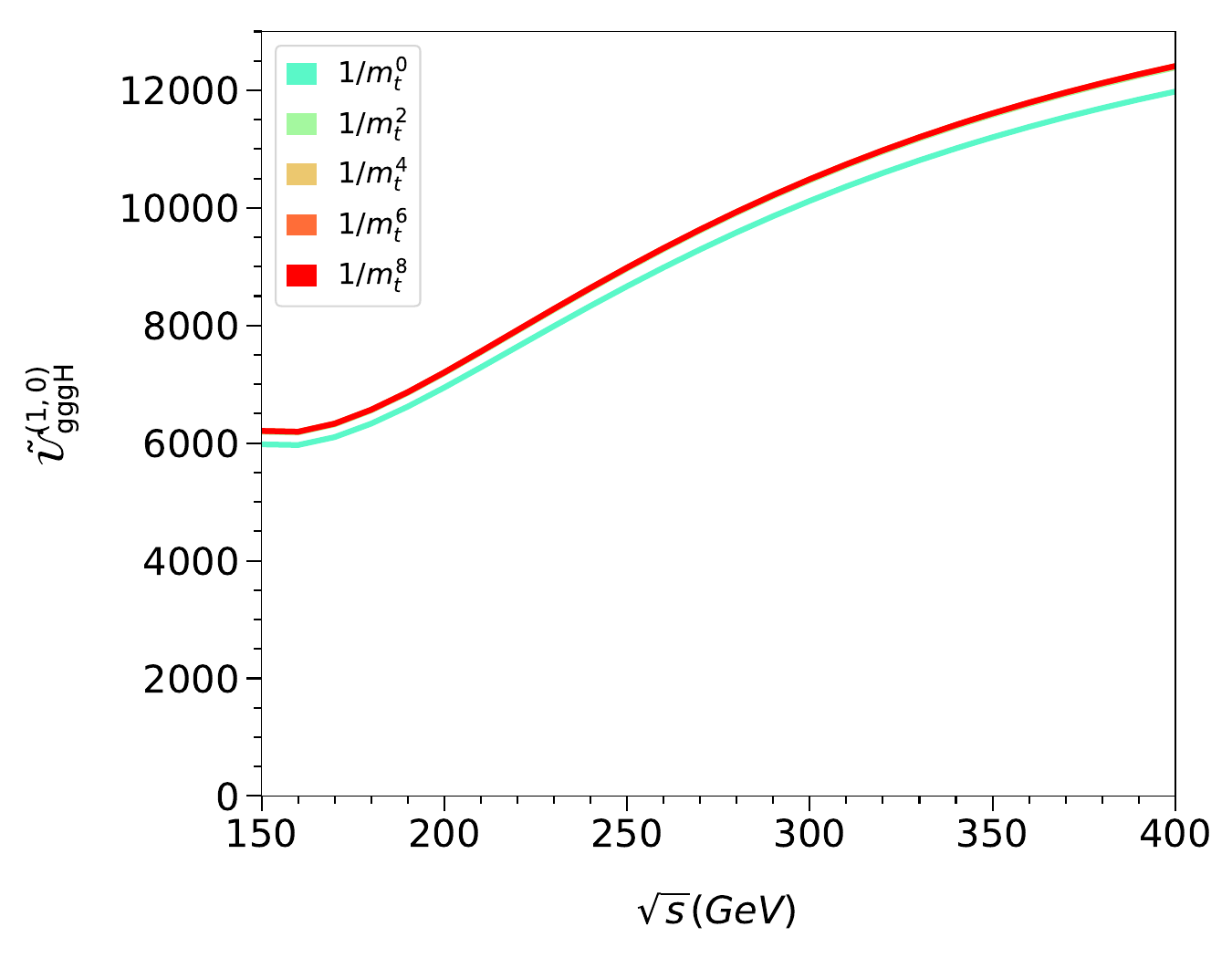} &
    \includegraphics[width=.45\textwidth]{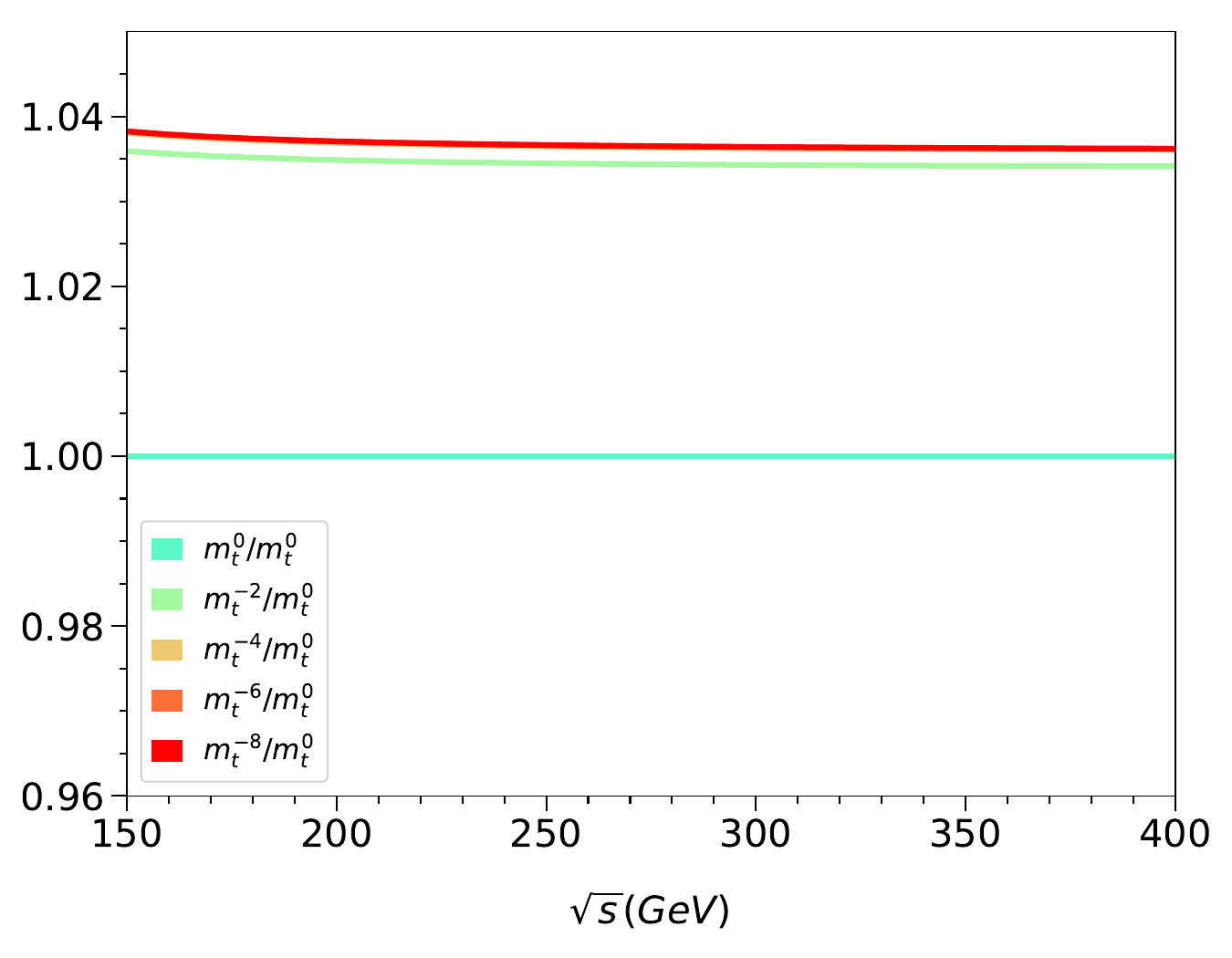}
  \end{tabular}
  \caption{\label{fig::gghg_2l_QCD}NLO QCD corrections to
    $\tilde{{\cal U}}_{\rm gggH}$ as a function of $\sqrt{s}$. For the
    construction of the squared matrix element the infra-red subtracted form
    factors (Eq.~\ref{eq::F_fin}) have been used.  The panel on the right
      shows the result normalized to the $m_t^0$ expansion term.}  
\end{figure}

In Fig.~\ref{fig::gghg_2l_QCD} we show the NLO QCD corrections to
${\cal U}_{\rm gggH}$ for $\rho_{p_T}=0.1$ as a function of $\sqrt{s}$. We
observe a rapid convergence, even beyond the top quark threshold (although the
expansion is not expected to produce the correct result in this region).
In fact, only the $1/m_t^2$ terms lead to a shift of a few percent; the
higher-order expansion terms are much smaller. This behaviour can be explained
by the dominance of the diagrams involving $ggH$ triangle contributions and
the suppression of the box-type Feynman diagrams.

%- }}}
%- {{{ Conclusions

\section{\label{sec::concl}Conclusions}

In this work we consider the gluon-fusion induced processes $gg\to HH$ and $gg\to gH$
and compute complete NLO electroweak corrections in the large top quark mass
limit and present results for the form factors up to order $1/m_t^{10}$ and
$1/m_t^{8}$, respectively.
We discuss the renormalization procedure in detail and compute all
counterterm contributions without assuming any mass hierarchy. Thus, this part
can also be applied to expansions in other kinematic limits or an exact
(numerical) calculation.

Partial electroweak results for $gg\to HH$ are already available in the
literature~\cite{Davies:2022ram,Muhlleitner:2022ijf}; in this work we
provide sub-leading terms in the large-$m_t$ expansion.

For $gg\to HH$ the expansion in $1/m_t$ does not show a convergent behaviour
in the physical region $2m_H \lesssim \sqrt{s} \lesssim 2m_t$.
We have demonstrated that this is due to
diagrams involving a cut through a $W$ boson and a top quark.
If these diagrams are omitted,
we observe reasonable convergence below
$\sqrt{s}\approx 330$~GeV.  Despite the limited applicability of the
large-$m_t$ expansion we believe that our results serve as reference for
future expansions in other kinematic regions or exact (numerical) calculations.
Despite the convergence issues, if we assume that the order of magnitude is at
least correct, in the large-$m_t$ region the electroweak contribution provides a
correction of a few tens of percent with respect to the leading order.

For the NLO electroweak corrections to $gg\to gH$ we observe very good convergence
below the top quark threshold.  In particular, for $\sqrt{s}<300$~GeV we can
provide precise predictions on the basis of an expansion which includes
corrections up to $1/m_t^8$. In this region the electroweak corrections are small,
below the percent level with respect to the leading order.

We also provide NLO QCD corrections for the four form factors needed for
$gg\to gH$ up to $1/m_t^8$. Here a rapid convergence is also observed
up to the top quark threshold.

%- }}}

%- {{{ Acknowledgements:

\section*{Acknowledgements}  

This research was supported by the Deutsche Forschungsgemeinschaft (DFG,
German Research Foundation) under grant 396021762 --- TRR 257 ``Particle
Physics Phenomenology after the Higgs Discovery'' and has received funding
from the European Research Council (ERC) under the European Union's Horizon
2020 research and innovation programme grant agreement 101019620 (ERC Advanced
Grant TOPUP).  The work of JD was supported by the Science and Technology
Facilities Council (STFC) under the Consolidated Grant ST/T00102X/1.  We thank
Christian Sturm for many useful discussions in connection to the
renormalization of the Higgs-gluon interactions. We also thank Martin Lang for
support in connection to the {\rm UFO} setup of {\rm tapir}.  We thank
Michael Spira for useful comments concerning Ref.~\cite{Muhlleitner:2022ijf}.
We have
used the program {\tt FeynGame}~\cite{Harlander:2020cyh} to draw the Feynman
diagrams.

%- }}}

\end{document}